# A Deterministic Algorithm for Pliable Index Coding

Linqi Song
Electrical Engineering Department
University of California, Los Angeles, USA
Email: songlinqi@ucla.edu

Christina Fragouli
Electrical Engineering Department
University of California, Los Angeles, USA
Email: christina.fragouli@ucla.edu

*Abstract*—Pliable index coding considers a server with $m$ messages, and $n$ clients where each has as side information a subset of the messages. We seek to minimize the number of transmissions the server should make, so that each client receives (any) one message she does not already have. Previous work has shown that the server can achieve this using $O(\log^2(n))$ transmissions and needs at least $\Omega(\log(n))$ transmissions in the worst case, but finding a code of optimal length is NP-hard. In this paper, we propose a deterministic algorithm that we prove achieves this upper bound, that is, in an order almost as the worst-case optimal code length. We also establish a connection between the pliable index coding problem and the minrank problem over a family of mixed matrices.

## I. INTRODUCTION

The conventional index coding problem, introduced in [1], considers a server with $m$ messages and $n$ clients. Each client has as side-information a subset of the messages and requires a specific message she does not have. The aim is to find an efficient way of sending the messages such that the number of transmissions is minimized. It has been shown that this problem is NP-hard, in the worst case we may require $\Omega(n)$ transmissions, and even for random graphs we will almost surely require $\Omega(\sqrt{n})$ transmissions [10].

Pliable index coding, introduced in [7], [8], still considers a server and $n$ clients with side information, but now assumes that the clients are pliable, and are happy to receive any one message they do not already have. For instance, when serving sale coupons inside a shopping mall, a client may not know in advance what are all the existing coupons, and is happy to receive any coupon she does not already have. Pliable index coding requires an exponentially smaller number of transmissions, in the worst case $O(\log^2(n))$ [7], [8]. Our previous results have shown that there exist instances requiring at least $\Omega(\log(n))$ transmissions, and thus the $O(\log^2(n))$ upper bound is almost tight [9]. These results imply that, if we realize that we need to solve a pliable index coding problem, as opposed to the conventional index coding problem, we can be exponentially more efficient in terms of the number of transmissions. However, the pliable index coding problem is NP-hard [7], and thus a natural question is, whether we can efficiently realize these benefits.

The main contribution of this paper is the design of a deterministic polynomial time algorithm for pliable index coding that always requires at most $O(\log^2(n))$ transmissions. This establishes that, although the problem is NP-hard, we can still in polynomial time achieve exponential benefits over index coding. To design this algorithm, we leverage an algebraic criterion for pliable index coding we have derived in [9]. We divide the transmissions into rounds, and in each round, we strategically group the messages into groups and use greedy transmission scheme to guarantee that a certain fraction of clients are satisfied in each round. Clearly our algorithm does not achieve the optimal code length, but still achieves an upper bound almost in the same order as the worst-case optimal code length. We derive theoretical bounds on the approximation ratio that measures how far away we are from the optimal exponential complexity algorithm. We also provide evaluation through simulations over random graphs and show that the proposed algorithm outperforms previously proposed heuristics [7] by up to 35% in some cases.

The paper makes two additional contributions: we prove that pliable index coding can be reduced to a min-rank problem over a family of matrices with constraints, as is also the case for index coding over a different family of matrices. We also provide an example to show that linear combining over the binary field is suboptimal for pliable index coding - we can get a better performance by enabling operations over larger finite fields.

## II. PROBLEM FORMULATION AND REPRESENTATION

We consider a broadcasting system with one server and $n$ clients. The server has $m$ messages, $b_1, b_2, \ldots, b_m$. We denote by $[m]$ and $[n]$ the sets $\{1, 2, \ldots, m\}$ and $\{1, 2, \ldots, n\}$. Each client $i$ has as side information a subset of messages, indexed by $S_i \subseteq [m]$, and requires a new message from the remaining subset of messages, called requirement subset and indexed by $R_i = [m] \backslash S_i$. The server makes broadcast transmissions (that may contain linear combinations of $b_1, b_2, \ldots, b_m$) over a noiseless broadcast channel. Each client then decodes the messages using her side-information and the received broadcast transmissions. In this paper, we restrict the encoding/decoding scheme to be linear and the messages to be in a finite field $\mathbf{F}_q$. In particular, we assume that the $k$-th transmission $x_k$ is a linear combination of $b_1, b_2, \ldots, b_m$, namely, $x_k = a_{k1}b_1 + a_{k2}b_2 + \ldots + a_{km}b_m$, where $b_j, x_k, a_{kj} \in \mathbf{F}_q$ and $a_{k1}, \ldots, a_{km}$ are the encoding coefficients. Therefore, we can interpret the number of transmissions, $K$, as the *code length* and the $K \times m$ coefficient matrix $\mathbf{A}$ with entries $a_{kj}$ as the *coding matrix*. In matrix form, we can write

$$\mathbf{x} = \mathbf{A}\mathbf{b}, \tag{1}$$

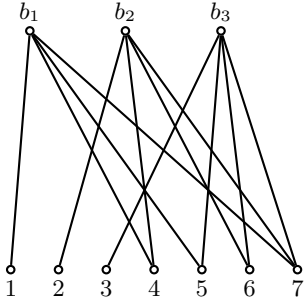

Fig. 1: A pliable index coding instance with $m = 3, n = 7$.

where $\boldsymbol{b}$ and $\boldsymbol{x}$ are the vectors of the original messages and encoded transmissions. Our goal is to construct the coding matrix $\boldsymbol{A}$, so that the code length $K$ is minimized.

Given $\boldsymbol{A}$, $\boldsymbol{x}$, and $\{b_j | j \in S_i\}$, the decoding process for client $i$ is to solve the linear equation (1) to get a unique solution of $b_{j^*}$, for some $j^* \in R_i$. Clearly, client $i$ can remove from the transmissions her side information messages, i.e., to recover $x_k^{(i)} = x_k - \sum_{j \in S_i} a_{kj} b_j$ from the $k$-th transmission. As a result, client $i$ only needs to solve the equations

$$\boldsymbol{A}_{R_i} \boldsymbol{b}_{R_i} = \boldsymbol{x}^{(i)}, \quad (2)$$

to retrieve any one message she does not have, where $\boldsymbol{A}_{R_i}$ is the sub-matrix of $\boldsymbol{A}$ with columns indexed by $R_i$; $\boldsymbol{b}_{R_i}$ is the message vector with elements indexed by $R_i$; and $\boldsymbol{x}^{(i)}$ is a $K$-dimensional column vector with element $x_k^{(i)}$.

### A. Bipartite graph representation

We can represent a pliable index coding problem using an undirected bipartite graph, where on one side we have a vertex corresponding to each of the $m$ messages, say $b_1$, ..., $b_m$ and on the other side one vertex corresponding to each client, $1, \ldots, n$ [7]. We connect with edges clients to the messages they *do not* have, i.e., client $i$ connects to the messages indexed by $R_i$. For instance, in the example in Fig. 1, $R_1 = \{1\}$ and $S_1 = \{2, 3\}$ for client 1; client 4 does not have (and would be happy to receive any of) $b_1$ and $b_2$. In this example, if the server transmits $x_1 = b_1 + b_2 + b_3$, $x_2 = b_2 + b_3$, and $x_3 = b_1 + b_2$, then we can write

$$\begin{bmatrix} 1 & 1 & 1 \\ 0 & 1 & 1 \\ 1 & 1 & 0 \end{bmatrix} \begin{bmatrix} b_1 \\ b_2 \\ b_3 \end{bmatrix} = \begin{bmatrix} x_1 \\ x_2 \\ x_3 \end{bmatrix}, \quad (3)$$

and the decoding process for client 4 is equivalent to solving the equations

$$\begin{bmatrix} 1 & 1 \\ 0 & 1 \\ 1 & 1 \end{bmatrix} \begin{bmatrix} b_1 \\ b_2 \end{bmatrix} = \begin{bmatrix} x_1 - b_3 \\ x_2 - b_3 \\ x_3 \end{bmatrix}. \quad (4)$$

### B. An algebraic criterion for pliable index coding

We here review an algebraic decodability criterion we introduced in [9]. Client $i$ needs to solve the linear equations (2) in order to recover a message she does not have. A key difference from classical decoding is that as long as (any) one variable $b_j$, $j \in R_i$ is recovered, the client $i$ is satisfied. Note that when solving the linear equations, there will in general exist multiple solutions; what we need is that all solutions for $\boldsymbol{b}_{R_i}$ give a unique value to $b_j$. Leveraging this observation, the following lemma checks whether client $i$ can recover message $b_j$ given a coding matrix $\boldsymbol{A}$. Here, we use $\boldsymbol{a}_j$ to denote the $j$-th column of matrix $\boldsymbol{A}$ and $\boldsymbol{A}_{R_i \setminus \{j\}}$ to denote a submatrix of $\boldsymbol{A}$ whose columns are indexed by $R_i$ other than $j$.

**Lemma 1.** *Client $i$ can uniquely decode message $b_j$, $j \in R_i$, if and only if*

$$\boldsymbol{a}_j \notin span\{\boldsymbol{A}_{R_i \setminus \{j\}}\}, \quad (5)$$

*where $span\{\boldsymbol{A}_{R_i \setminus \{j\}}\} = \{\sum_{l \in R_i \setminus \{j\}} \lambda_l \boldsymbol{a}_l | \lambda_l \in \boldsymbol{F}_q\}$ is the linear space spanned by columns of $\boldsymbol{A}_{R_i \setminus \{j\}}$.*

*Proof.* : See [9]. □

For example, for the instance in Fig. 1, we have $R_4 = 1, 2$, $\boldsymbol{a}_1 \notin span\{\boldsymbol{a}_2\}$, and $\boldsymbol{a}_2 \notin span\{\boldsymbol{a}_1\}$, so client 4 can decode $b_1$ and $b_2$. This is indeed possible because client 4 can decode $b_2$ by $b_2 = x_2 - b_3$ and $b_1$ by $b_1 = x_3 - b_2$.

## III. DETERMINISTIC ALGORITHM

We propose a greedy algorithm to realize the pliable index coding which runs in rounds. In each round, the algorithm splits the messages into at most $\log n$ groups and makes two transmissions per group. We prove that our greedy algorithm manages to satisfy at least $1/3$ of the clients with $2 \log n$ transmissions. Thus, repeating $O(\log n)$ rounds, we can satisfy all clients using $O(\log^2(n))$ transmissions. We start by describing the greedy algorithm, and then prove its performance bounds as well as approximation ratio bounds.

### A. Greedy Algorithm Description

We here give an overview of the binary field greedy (Bin-Greedy) algorithm in Alg. 1. The algorithm uses the bipartite graph representation of pliable index coding and operates in rounds. Each round has two phases: the sorting phase and the greedy transmission phase. In the sorting phase, the algorithm sorts the message vertices in a decreasing order in terms of their *effective degrees* that we will define later, and divides the messages into $\log n$ groups. In the transmission phase, it selects linear combinations and makes two transmissions per message group, thus in total $2 \log n$ transmissions.

- **Effective degree** and **effective clients**: given a particular order of the message vertices $\pi = (j_1, j_2, \ldots, j_m)$, the effective degree of message $b_{j_l}$ is defined as the number of $b_{j_l}$'s neighbors who do not connect with message $b_{j'}$, for any $j' = j_1, j_2, \ldots, j_{l-1}$. These neighbors counting for $b_{j_l}$'s effective degree are called effective clients of $b_{j_l}$. Let us denote by $N[j]$ the set of neighbors of message $b_j$ and by $N[j_1, j_2, \ldots, j_{l-1}]$ the set $N[j_1] \cup N[j_2] \cup \ldots N[j_{l-1}]$.

Formally, the effective clients of message $b_{j_l}$ are defined as $N_\pi^\dagger[j_l] = N[j_l]\setminus N[j_1, j_2, \ldots, j_{l-1}]$ with respect to the message order $\pi$. Correspondingly, the effective degree of message $b_{j_l}$ is defined as $d_\pi^\dagger[j_l] = |N_\pi^\dagger[j_l]|$ with respect to $\pi$.

Note that the effective degree and effective clients for a message $b_j$ may vary when we change the order of the message vertices. We will omit the subscript $\pi$ when it is clear from the contexts. In our example in Fig. 1, given a message order $b_1, b_2, b_3$, the effective degrees and clients are $d^\dagger[1] = 4, N^\dagger[1] = \{1, 4, 5, 7\}$, $d^\dagger[2] = 2, N^\dagger[2] = \{2, 6\}$, and $d^\dagger[3] = 1, N^\dagger[3] = \{3\}$.

*Sorting Phase*: in the following, we will describe how we sort the messages into a desired order.

• Step 1: We denote the original bipartite graph by $G_1$. Find a message vertex $j_1$ with the maximum degree (number of neighbors) in $G_1$, with ties broken arbitrarily. Thus we have $|N_{G_1}[j_1]| \geq |N_{G_1}[j]|$ for all $j \in [m]\setminus\{j_1\}$, where $N_{G_1}[j] = N[j]$.

• Step 2: Consider the subgraph induced by message vertices $[m]\setminus\{j_1\}$ and client vertices $[n]\setminus N[j_1]$, denoted by $G_2$. Find a message vertex $j_2$ with maximum degree in the subgraph $G_2$, with ties broken arbitrarily. That is, we have $|N_{G_2}[j_2]| \geq |N_{G_2}[j]|$ for all $j \in [m]\setminus\{j_1, j_2\}$, where $N_{G_2}[j] = N[j] \cap V(G_2) = N[j]\setminus N[j_1]$. Here $V(G_2)$ are the vertex set of subgraph $G_2$.

In general, we can repeat the sorting process for step $l = 1, 2, \ldots, m$.

• Step $l$: Consider the subgraph induced by message vertices $[m]\setminus\{j_1, j_2, \ldots, j_{l-1}\}$ and client vertices $[n]\setminus(N[j_1, j_2, \ldots, j_{l-1}]$, denoted by $G_l$. Find a message vertex $j_l$ with maximum degree in the subgraph $G_l$, with ties broken arbitrarily. That is, we have $|N_{G_l}[j_l]| \geq |N_{G_l}[j]|$ for all $j \in [m]\setminus\{j_1, j_2, \ldots, j_l\}$, where $N_{G_l}[j] = N[j] \cap V(G_l) = N[j]\setminus N[j_1, j_2, \ldots, j_{l-1}]$. Here $V(G_l)$ are the vertex set of subgraph $G_l$.

From the above sorting process, we notice that the effective degrees are $|N[j_1]|$ for message $j_1$, $|N[j_2]\setminus N[j_1]|$ for message $j_2$, …, $|N[j_l]\setminus N[j_1, j_2, \ldots, j_{l-1}]|$ for $j_l$, etc.

We divide the message vertices into $\log(n)$ groups, $\mathcal{M}_1, \mathcal{M}_2, \ldots, \mathcal{M}_{\log(n)}$ based on their effective degrees with respect to the above order, such that for message vertex $j \in \mathcal{M}_s$, the effective degree satisfies $n/2^{s-1} \geq d^\dagger[j] > n/2^s$.

According to the above sorting and grouping process, we have the following property for the message $j$ in group $\mathcal{M}_s$:

$$d^\dagger[j] > n/2^s \triangleq \frac{d}{2}, \text{ and } |N[j] \cap \mathcal{N}_s| \leq n/2^{s-1} \triangleq d, \quad (6)$$

where $\mathcal{N}_s = \cup_{j' \in \mathcal{M}_s} N^\dagger[j']$. The second part holds because if $|N[j] \cap \mathcal{N}_s| > d$, the message $j'$ will be assigned in a group less than $s$ in the sorting and grouping process.

Applying the sorting procedure to our example in Fig. 1, we can find one possible sorting order $b_1, b_2, b_3$ and groups $\mathcal{M}_1 = \{1\}, \mathcal{M}_2 = \{2\}, \mathcal{M}_3 = \{3\}$.

*Transmission Phase*: We make two transmissions for each message group $\mathcal{M}_s$, using a sub-coding matrix with 2 rows (one for each transmission). We sequentially create this matrix by visiting each of the vertices in the group $\mathcal{M}_s$, according to the sorting order, and adding for each vertex one column to the sub-coding matrix (we refer to this column as the coding vector associated with this message). We select each column to be one in the set $\{(1, 0)^T, (0, 1)^T, (1, 1)^T\}$, by greedily evaluating and selecting the one that can satisfy as many clients in $\mathcal{N}_s$ as possible. In our example in Fig. 1, we can construct a coding matrix:

$$A = \begin{bmatrix} 1 & 0 & 0 \\ 0 & 0 & 0 \\ \hline 0 & 1 & 0 \\ 0 & 0 & 0 \\ \hline 0 & 0 & 1 \\ 0 & 0 & 0 \end{bmatrix}, \quad (7)$$

where every two rows represent the transmissions for a group.

---

**Algorithm 1** Binary Field Greedy Algorithm (BinGreedy)

1: **Initialization**: Set $\mathcal{N} = [n]$.
2: **while** $\mathcal{N} \neq \emptyset$ **do**
3:    **Sorting and grouping of message vertices**:
4:    Set $\mathcal{N}_{temp} = \mathcal{N}, \mathcal{M}_{temp} = [m]$.
5:    **for** $j = 1 : m$ **do**
6:      Find the message $j' \in \mathcal{M}_{temp}$ having the maximum number of neighbors in $\mathcal{N}_{temp}$, with ties broken arbitrarily.
7:      Put message $j'$ in the $j$-th position.
8:      Remove $j'$ from $\mathcal{M}_{temp}$ and all its neighbors from $\mathcal{N}_{temp}$.
9:    **end for**
10:   Group messages into $\mathcal{M}_1, \mathcal{M}_2, \ldots, \mathcal{M}_{\log(n)}$ message groups based on their effective degrees.
11:   **Greedy coding**:
12:   **for** $s = 1 : \log(n)$ **do**
13:      **Initialization**: Set $\mathcal{N}_s = \cup_{j \in \mathcal{M}_s} N^\dagger[j]$ (effective clients neighboring to $\mathcal{M}_s$), $SAT = \emptyset$ and $UNSAT = \emptyset$.
14:      **for** $j = 1 : |\mathcal{M}_s|$ **do**
15:         Assign a coding vector from the set $\{(1, 0)^T, (0, 1)^T, (1, 1)^T\}$ to the $j$-th message in $\mathcal{M}_s$, such that the maximum number of clients in $\{i \in SAT | i \text{ is connected with } j\}$ can still be satisfied, with ties broken arbitrarily.
16:         Move from $SAT$ to $UNSAT$ these unsatisfied clients in $\{i \in SAT | i \text{ is connected with } j\}$.
17:         Add clients in $N^\dagger[j]$ to $SAT$.
18:      **end for**
19:      Set coding vectors to be $(0, 0)^T$ corresponding to messages in groups other than $s$.
20:      Remove clients in $SAT$ from $\mathcal{N}$ and their associated edges.
21:   **end for**
22: **end while**

*B. Algorithm performance*

The following lemma states that using the greedy coding scheme, at least a fraction $1/3$ of the effective clients connecting with $\mathcal{M}_s$, i.e., $\mathcal{N}_s$, can be satisfied.

**Lemma 2.** *In Alg. 1, the greedy coding scheme can satisfy a fraction of at least $1/3$ of the effective clients $\mathcal{N}_s$.*

*Proof.* In this proof, we restrict to the subgraph induced by vertices $\mathcal{M}_s \cup \mathcal{N}_s$ corresponding to the transmissions of groups $\mathcal{M}_s$ in a certain round. In each step, we sequentially assign to a message a coding vector in a greedy manner and try to dynamically evaluate whether each client is satisfied or not up to current step (by only considering messages visited up to now and disregarding all unvisited messages). We first observe that when we assign to the first message a coding vector $(1,0)^T, (0,1)^T$, or $(1,1)^T$, all clients connecting to this message will be satisfied. To capture this, we define two dynamic sets, $SAT$ and $UNSAT$. Assume that up to some step, the algorithm visits some messages and assigned corresponding coding vectors. The first set, $SAT$, collects the clients connecting to messages that have already been visited, and are satisfied by the current assignment of coding vectors according to the criterion in Lemma 1, i.e., for each of these clients, $i$, we consider the $r$ coding vectors corresponding to messages connecting with $i$ and visited by the algorithm so far, $\alpha_1, \alpha_2, \ldots, \alpha_r$, there exist one coding vector $\alpha_j$ ($1 \leq j \leq r$) that does not in the span of the remaining coding vectors $\alpha_j \notin \text{span}\{\alpha_1, \ldots, \alpha_{j-1}, \alpha_{j+1}, \ldots, \alpha_r\}$. The second set, $UNSAT$, collects the set of clients that are associated with messages already visited by the algorithm and cannot be satisfied by current coding vector assignments. Note that there also exist clients that are in neither of these groups, as the messages they are associated with have not yet been visited.

Initially, both $SAT$ and $UNSAT$ are empty. We gradually add clients from $\mathcal{N}_s$ into these two sets as we go through the messages and assign coding vectors. Our first step is to add all $N^\dagger[1]$ to $SAT$, since any non-zero vector satisfies the decoding criterion for only one message. We try to analyse the algorithm's behavior at some point as some satisfied clients may become unsatisfied when we assign more coding vectors to messages connecting with these clients. For example, a client is connected with 3 messages, 2 of which are visited and assigned coding vectors $(1,0)^T, (0,1)^T$, so this client is satisfied up to now. However, when the algorithm visits the third message and assigns a coding vector $(1,1)^T$, this client becomes unsatisfied as the decoding criterion does not hold at this time.

Next, we want to show that each step, the number of clients who are moved from $SAT$ to $UNSAT$ is at most $\frac{d}{3}$. Notice that when we assign a coding vector $(1,0)^T, (0,1)^T$, or $(1,1)^T$ to message $j$, only clients connecting with message $j$ can be affected. We list possibilities for all $t$ clients connecting with $j$ and are satisfied (in $SAT$) before step $j$:

• Case 1: Assume there are $t_1$ clients who connect with previously visited messages that are assigned a coding vector $(1,0)^T$ and several (can be 0) coding vectors $(0,1)^T$. In this case, these clients can decode a new message corresponding to the coding vector $(1,0)^T$ since $(1,0)^T$ does not belong in the span of $(0,1)^T$. Similarly,

• Case 2: Assume $t_2$ clients are satisfied by one $(1,0)^T$ and several $(1,1)^T$.

• Case 3: Assume $t_3$ clients are satisfied by one $(0,1)^T$ and several $(1,0)^T$.

• Case 4: Assume $t_4$ clients are satisfied by one $(0,1)^T$ and several $(1,1)^T$.

• Case 5: Assume $t_5$ clients are satisfied by one $(1,1)^T$ and several $(0,1)^T$.

• Case 6: Assume $t_6$ clients are satisfied by one $(1,1)^T$ and several $(1,0)^T$.

When we assign a coding vector $(1,0)^T$ to message $j$, then the $t_3 + t_6$ clients can still be satisfied according to our decoding criterion of Lemma 1. Similarly, if we assign a coding vector $(0,1)^T$ or $(1,1)^T$ to message $j$, then the $t_1 + t_5$ or $t_2 + t_4$ clients can still be satisfied.

Note that $t_1 + t_2 + t_3 + t_4 + t_5 + t_6 \geq t$ as there may be overlap among the 6 different cases (e.g., a client is satisfied by one $(1,0)^T$ and one $(0,1)^T$, so she is counted twice in both Case 1 and Case 3). Hence, at least one of $t_3 + t_6$, $t_1 + t_5$, $t_2 + t_4$ should be no less than $t/3$; our greedy algorithm will move at most $2t/3$ clients from $SAT$ to $UNSAT$. According to the property of our sorting and grouping in eq. (6), the number of $j$'s neighbors who are connected with previously visited message is at most $d - d^\dagger[j] < d/2$, and furthermore the number of $j$'s neighbors in set $SAT$ is a subset of these neighbors, resulting in $t < d/2$. So at most $d/3$ clients will be moved from $SAT$ to $UNSAT$ in each step.

On the other hand, we observe that for message $j$'s effective clients ($j$'s neighbors who are not connected with previously visited messages), any assignment of vectors $(1,0)^T, (0,1)^T$, or $(1,1)^T$ can satisfy them according to the decoding criterion. Hence, at least $d^\dagger[j] > d/2$ new clients are added to the $SAT$ set. Repeating the assigning steps, we can see that at most $(d/3)/(d/2) = 2/3$ clients in $\mathcal{N}_s$ cannot be satisfied by this coding scheme. $\square$

Therefore, we have the following theorem to show the performance of the greedy algorithm.

**Theorem 1.** *For the BinGreedy algorithm, Alg. 1, the number of required transmissions is at most $O(\log^2(n))$.*

*Proof.* From Lemma 2, in each round, we have at most $\log(n)$ groups and $2\log(n)$ transmissions such that a fraction of at least $1/3$ clients are satisfied. This can be repeated for at most $O(\log(n))$ times, resulting in an upper bound of $O(\log^2(n))$ for the greedy algorithm. $\square$

From the construction of our greedy algorithm, we notice that this algorithm runs in polynomial time $O(nm^2 \log(n))$. Because there are at most $O(\log(n))$ rounds; for each round, the sorting and grouping process take time $O(nm^2)$; and the greedy coding in each round takes time $O(mn)$.

*C. Approximation ratio*

We here theoretically compare how our algorithm performs as compared to the optimal (exponential complexity) algorithm. The approximation ratio $\alpha(n)$ is defined as the maximum ratio of the code length of our algorithm and that of the optimal algorithm among all instances of client size $n$, i.e.,

$$\alpha(n) = \max_{I_n} \frac{BinGreedy(I_n)}{OPT(I_n)}, \quad (8)$$

where $I_n$ is any instance with client size $n$; $BinGreedy(I_n)$ is the code length of our algorithm for instance $I_n$; and $OPT(I_n)$ is the optimal code length for instance $I_n$. Our next theorem states that the approximation ratio is bounded by $\Omega(\log\log(n))$ and $O(\log^2(n))$.

**Theorem 2.** *The approximation ratio of our greedy algorithm $\alpha(n)$ satisfies $\Omega(\log\log(n)) \leq \alpha(n) \leq O(\log^2(n))$, unless $NP \subseteq BPTIME(n^{O(\log\log(n))})$.*

The proof is provided in Appendix B. The proof uses a gap reduction from an NP-hard problem, the minimum representation problem, and shows that it is hard to approximate our pliable index coding problem using a code of length less than $\Omega(\log\log(n))$.

## IV. DISCUSSION

The main intuition behind the algorithm is the following. The effective degree captures which clients we are attempting to satisfy by transmitting a specific message. As mentioned, we will call the clients associated with the effective degree of the messages in a group as effective clients (these are the clients to be satisfied by this group of messages). If we create a group of message that have approximately the same effective degree, we can find a coding scheme with two transmissions satisfies at least a constant fraction of these clients. The proposed algorithm could potentially achieve a better performance by making for instance more than two transmissions per group (a variable number of transmissions), or by using linear combining over larger finite fields (our algorithm only uses the binary field). Yet, it is not possible to design an algorithm that has a better worst-case-performance, since we achieve the same worst-case-performance as the optimal (exponential in complexity) algorithm. Next, we first give an example that shows we could have a better performance by using a higher finite field and we make a connection between the pliable index coding problem and a min-rank problem. This connection shows that our algorithm can also be used for solving this set of minrank problems.

*Field size:* One interesting question is whether binary field is enough to achieve the optimal code length. We show through a counter example that a binary code may not be sufficient. Consider the following instance with $m = 4$ and $n = 10$:

- $R_1 = \{1\}, R_2 = \{2\}, R_3 = \{3\}, R_4 = \{4\}, R_5 = \{1,2\}, R_6 = \{1,3\}, R_7 = \{1,4\}, R_8 = \{2,3\}, R_9 = \{2,4\}, R_{10} = \{3,4\}$.

This instance contains clients with requirement sets of all 1 message and 2 message subsets. We can easily see that the optimal code length is 2, e.g., $b_1+b_2+b_4$ and $b_2+b_3+2b_4$, in $\mathbf{F}_3$. However, we cannot find a binary code of length 2, because we have all 1 message and 2 messages requirement sets, requiring $\boldsymbol{a}_j \neq (0,0)^T$, for $j = 1,2,3,4$ and $\boldsymbol{a}_j \neq \boldsymbol{a}_{j'}$, for $j \neq j'$. But, we have only 3 non-zero vectors $(1,0)^T, (0,1)^T, (1,1)^T$. It is not possible to assign these 3 non-zero vectors to $\boldsymbol{a}_j$ such that any 2 of them satisfy all clients. In general, we need at least field size $m-1$ in order to achieve the optimal code length [1].

*Minrank:* In index coding, the optimal linear code length is characterized by a term minrank, which is the minimum rank of a mixed matrix (some of whose elements are to be determined) associated with the requirement graph [2]. In a similar way, we can characterize the pliable index coding problem using the minimum rank of a mixed matrix associated with the bipartite requirement graph.

We say that a matrix $\boldsymbol{G} \in \mathbf{F}_q^{n \times m}$ fits the pliable index coding problem $(m, n, \{R_i\}_{i \in [n]})$ if in the $i$-th row ($\forall i \in [n]$),
1) for $j \in R_i$, there exists one and only one $j^* \in R_i$, such that $g_{ij^*} = 1$, and other $g_{ij} = 0$ for any $j \in R_i \setminus \{j^*\}$;
2) for $j \in S_i$, $g_{ij}$ can be any element in $\mathbf{F}_q$.

Let us denote by $\mathcal{G}$ the set of all matrices fitting the pliable index coding problem $(m, n, \{R_i\}_{i \in [n]})$, and by minrank$(\mathcal{G})$ the minimum rank among all the fitted matrices $\boldsymbol{G} \in \mathcal{G}$. In other words, minrank$(\mathcal{G}) = \min_{\boldsymbol{G} \in \mathcal{G}} \text{rank}(\boldsymbol{G})$, where rank$(\boldsymbol{G})$ denote the rank of matrix $\boldsymbol{G}$. We have the following theorem to characterize the optimal coding length:

**Theorem 3.** *The optimal linear code length of the pliable index coding instance $(m, n, \{R_i\}_{i \in [n]})$ is minrank$(\boldsymbol{G})$.*

The proof is provided in Appendix A. The idea of the proof is to show that for a client $i$, there exists a linear combination of row vectors of the coding matrix such that there will be exactly one 1 and $|R_i| - 1$ 0s for elements indexed by the requirement set $R_i$.

## V. NUMERICAL RESULTS

We compare the performance of our proposed algorithm BinGreedy, with the randomized algorithm proposed in [7], which is the state-of-the art alternative currently proposed. The average performance with respect to the random code realization of this randomized algorithm is upper bounded by $O(\log^2(n))$. The randomized algorithm works as follows [7]: it splits the clients into $\log(n)$ bins based on their degrees, and makes transmissions for each set of clients separately. For bin $s$, the degrees of the clients is between $n/2^s$ and $n/2^{s-1}$, and a code for one transmission is generated by assigning each bit independently to be 1 with probability $2^s/n$ and 0 otherwise. Transmissions are repeated until all clients in the bin are satisfied. In contrast, BinGreedy (described in Section III) splits the messages into groups (not the clients), according to their effective degree (not their degree), and makes always two transmissions per group.

---
[1]See Appendix C for a discussion of this.

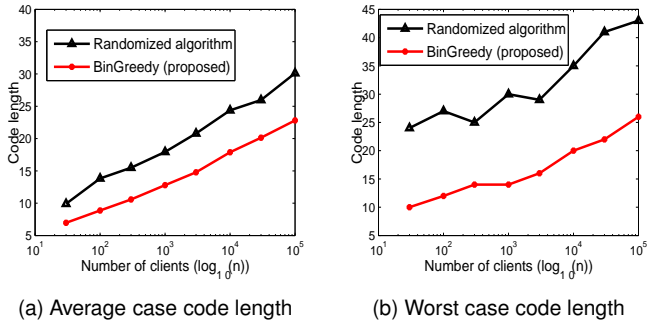

Fig. 2: Comparison of algorithm performances in terms of average case and worst case code length, achieved by BinGreedy and randomized algorithms.

We evaluate the performances of the two algorithms in terms of the code length (or number of transmissions) as the number of clients and number of messages vary. By setting the number of messages $m$ to be $n^{0.75}$, we numerically investigate relationship of the code length and the number of clients $n$ in the system. We randomly generate pliable index coding instance bipartite graph by connecting each client and each message with probability $0.3$. For each $n$, we randomly generate $100$ instances for simulation.

Fig. 2 shows the average case code length and worst case code length for each value of $n$. Note that the horizontal axis is in the scale of $\log_{10}(n)$. We can see that in the average case (averaged over $100$ instances for the same $n$), the proposed BinGreedy algorithm outperforms the randomized algorithm by $20\%$-$35\%$ in terms of code length; in the worst case, the proposed BinGreedy algorithm outperforms the randomized algorithm by $40\%$-$60\%$ in terms of code length.

Due to the randomization property of the randomized algorithm, different realizations may differ significantly. So we expect our proposed deterministic algorithm to be more robust than the randomized algorithm. As expected, we can see from Fig. 2 that the worst-case to average-case ratio ($n \geq 100$) is $1.4$-$2.0$ for the randomized algorithm and is $1.1$-$1.35$ for our proposed algorithm.

## VI. CONCLUSION

Although finding the optimal pliable index code length is NP-hard, we have proposed a deterministic algorithm to guarantee that the code length is at most $O(\log^2(n))$.

## APPENDIX A
## PROOF OF THE THEOREM ABOUT OPTIMAL CODE LENGTH

First, let us prove that a linear code with length $K = \text{minrank}(\mathcal{G})$ exists. Assume that a fitted matrix $\boldsymbol{G}$ achieves rank $K$. Without loss of generality, let us also assume that the first $K$ rows of $\boldsymbol{G}$ are linearly independent. For the encoding process, we define the coding matrix $\boldsymbol{A}$ to be the first $K$ rows of $\boldsymbol{G}$. For matrix $\boldsymbol{G}$, there is one and only one $j^* \in R_i$, such that $g_{ij^*} = 1$, and other $g_{ij} = 0$ for $j \in R_i \setminus \{j^*\}$; so that column $\boldsymbol{g}_{j^*}$ cannot be expressed as a linear combination of $\{\boldsymbol{g}_j\}_{j \in R_i \setminus \{j^*\}}$. Since all the rows of $\boldsymbol{G}$ are linear combinations of the first $K$ rows, column $\boldsymbol{a}_{j^*}$ cannot be expressed as a linear combination of $\{\boldsymbol{a}_j\}_{j \in R_i \setminus \{j^*\}}$ either. As a result, the decodable criterion holds for client $i$ and message $j^*$ can be decoded by client $i$.

Next, let us prove that for any linear code with a $K \times m$ coding matrix $\boldsymbol{A}$ in filed $\boldsymbol{F}_q$ has a code length $K \geq \text{minrank}(\mathcal{G})$. We show that using the coding matrix $\boldsymbol{A}$, we can build a matrix $\boldsymbol{G} \in \boldsymbol{F}_q^{n \times m}$ with rank at most $K$ that fits the index coding problem. To show this, we use the following claim.

**Claim 1**: If for client $i$, the message $j^*$ can be decoded, then the row vector $e_{j^*}^T$ is in the span of $\{\boldsymbol{\alpha}_l^T : l \in [K]\} \cup \{e_j^T : j \in S_i\}$, where $e_j^T$ is a row vector with all 0s, except a 1 in the $j$-th position and $\boldsymbol{\alpha}_l^T$ represents the $l$-th row of matrix $\boldsymbol{A}$.

This claim shows that $e_{j^*}^T$ is in the span of the union of row space of $\boldsymbol{A}$ and the side-information space. The proof of this claim can be found in [2].

For each client $i$, the claim states that $e_{j^*}^T = \sum_{l=1}^{K} \lambda_l \boldsymbol{\alpha}_l^T + \sum_{j \in S_i} \mu_j e_j^T$ for some $\lambda_l, \mu_j$ in field $\boldsymbol{F}_q$. To construct $\boldsymbol{G}$, we define the $i$-th row of $\boldsymbol{G}$, $\gamma_i^T$, to be the linear combination $\sum_{l=1}^{K} \lambda_l \boldsymbol{\alpha}_l^T$. Or equivalently, we have $\gamma_i^T = \sum_{l=1}^{K} \lambda_l \boldsymbol{\alpha}_l^T = e_{j^*}^T - \sum_{j \in S_i} \mu_j e_j^T$. This shows that $\gamma_i^T$ has value 1 at position $j^*$, $-\mu_j$ at position $j \in S_i$, and 0 at positions indexed by $R_i \setminus \{j^*\}$.

Therefore, we have shown that $K \geq \text{rank}(\boldsymbol{G}) \geq \text{minrank}(\mathcal{G})$.

## APPENDIX B
## PROOF OF THE THEOREM ABOUT APPROXIMATION RATIO

For the upper bound part, we can directly get it from Theorem that the code length is upper bounded by $O(\log^2(n))$.

For the lower bound part, we use the same method as in paper [11] with a slight modification.

The basic idea of the proof is to use a gap reduction from the minimum representation problem (MIN-REP) to show the approximation ratio. We use the same reduction from a minimum representation problem instance as in [11]. Then we have the following property of our pliable index coding problem:

- (1) Yes instance for MIN-REP: we have a broadcast scheme for our corresponding our pliable index coding instance such that the code length is as most 2.
- (2) $\log^{10}$-No instances for MIN-REP: with probability at least $1 - 2^{-\Omega(n^{5/9})}$, our pliable index code length is bounded by $O(\log \log(n))$.

The Property (1) can be directly derived from the construction of the reduction process. While for Property (2), we observe that if we use our pliable index code $[\alpha_1, \alpha_2, \ldots, \alpha_{O(\log \log(n))}]$ and their linear combinations to generate a new set of code $\beta_1, \beta_2, \ldots, \beta_B$. Then this set of code can be used as a schedule in the radio broadcast problem in [11]. However, the length of the new set of code $B = O(\log(n))$ for a fixed filed $\mathbb{F}_q$. From Corollary 6.4 in [11], we can get the Property (2).

So that this reduction transforms a Yes-instance of MIN-REP into an instance with constant optimal pliable index code length; and transforms a $\log^1 0$-No instance of MIN-REP into an instance of optimal code length $O(\log \log(n))$ with high probability. Since it is hard to distinguish the Yes-instance and $\log^{10}$-No instance of MIN-REP, our result follows from the PCP theorem.

## APPENDIX C
## DISCUSSION ABOUT FIELD SIZE

In this section, we discuss about the field size required for an instance with $m$ messages and $n$ clients. We consider an instance with $m$ messages and $n = m + \binom{m}{2}$ clients, where the clients have all 1-element and 2-element requirement sets. Namely, the clients' requirement sets are $\{j\}$ and $\{j_1, j_2\}$, for any $j \in [m]$ and $j_1, j_2 \in [m]$.

In this case, we show that the field size needs to be at least $m - 1$ in order to achieve an optimal code length. According to network coding theorem, we know that the optimal code length for such an instance will be 2 where we can assign coding vectors of length 2 to messages such that any two of the coding vectors are linearly independent.

Assume we use finite field $\boldsymbol{F}_q$ to realize the coding. According to our decoding criterion, we need every coding vector to be nonzero and any pair of the coding vectors to be linearly independent.

- If the coding vector contains 0, then there will be 2 of them: $(1,0)^T$ and $(0,1)^T$ since any other vector in the form of $(x,0)^T$ $(x \in \boldsymbol{F}_q)$ is linearly dependent with $(1,0)^T$ and similarly, $(0,x)^T$ $(x \in \boldsymbol{F}_q)$ is linearly dependent with $(0,1)^T$.
- If the coding vector is in the form $(x,y)^T$, $x,y \in \boldsymbol{F}_q, x,y \neq 0$, then there are in total $(q-1)^2$ such vectors. However, $(x,y)^T$ is linearly dependent with $z(x,y)^T$, for $z \in \boldsymbol{F}_q$. There are in total $(q-1)$ distinct $z(x,y)^T$ vectors, so the total number of pair-wise independent vectors is $(q-1)^2/(q-1) = (q-1)$.

Therefore, we need $2 + (q-1) \geq m$ in order to satisfy these clients, resulting in $q \geq m - 1$.